\documentclass[conference]{IEEEtran}
\IEEEoverridecommandlockouts
\usepackage{cite}
\usepackage{amsmath,amssymb,amsfonts}
\usepackage{algorithmic}
\usepackage{graphicx}
\usepackage{textcomp}
\usepackage{xcolor}
\usepackage{tabularx,adjustbox,booktabs}
\usepackage{colortbl}
\usepackage{multirow}
\usepackage{caption}
\usepackage{subcaption}
\usepackage{mathtools}

\newcommand\clearrow{\global\let\rowmac\relax}
\clearrow

\def\BibTeX{{\rm B\kern-.05em{\sc i\kern-.025em b}\kern-.08em
    T\kern-.1667em\lower.7ex\hbox{E}\kern-.125emX}}
\begin{document}

\title{SpotHitPy: A Study For ML-Based Song Hit Prediction Using Spotify\\}

\author{\IEEEauthorblockN{Ioannis Dimolitsas\IEEEauthorrefmark{1}, Spyridon Kantarelis\IEEEauthorrefmark{2}and Afroditi Fouka\IEEEauthorrefmark{3}}
\IEEEauthorblockA{\textit{School of Electrical and Computer Engineering,} \\
\textit{National Technical University of Athens,}
Athens, Greece \\
Email: \IEEEauthorrefmark{1}jdimol@netmode.ntua.gr,
\IEEEauthorrefmark{2}spyroskanta@ails.ece.ntua.gr,
\IEEEauthorrefmark{3}afroditifouka@mail.ntua.gr} \\ 
% \IEEEauthorrefmark{4}{Corresponding author.}
% \textit{{jdimol@netmode.ntua.gr, spyroskanta@ails.ece.ntua.gr, afroditifouka@mail.ntua.gr}}}
}

% \author{\IEEEauthorblockN{Ioannis Dimolitsas\\\textit{jdimol@netmode.ntua.gr}}

% \and
% \IEEEauthorblockN{Spyros Kantarelis\\\textit{spyroskanta@ails.ece.ntua.gr}}
% \and
% \IEEEauthorblockN{Afroditi Fouka\\\textit{afroditifouka@mail.ntua.gr}}
% }

\maketitle

\begin{abstract}
In this study, we approached the Hit Song Prediction problem, which aims to predict which songs will become Billboard hits. We gathered
a dataset of nearly 18500 hit and non-hit songs and extracted their audio features using the Spotify Web API. We test four machine-learning models on our dataset. We were able to predict the Billboard success of a song with approximately 86\% accuracy. The most succesful algorithms were Random Forest and Support Vector Machine.
\end{abstract}


\section{Introduction}
Hit Song Science (HSS) is an active research topic in Music Information Retrieval (MIR). The main focus of this topic
is to predict whether a song will become a hit or not. Hit prediction is therefore useful to musicians, labels, and music vendors considering that popular songs address to a big audience. Hit songs help labels and music vendors to increase their profits, and artists to share their message with a broad audience. Our approach to this topic relies on the assumption that hit songs are similar with respect to their audio features. We gathered a dataset of hit and non-hit songs and their features using the Spotify Web API\footnote{https://developer.spotify.com/documentation/web-api/}. We then use machine learning methods and algorithms to predict if a song is a hit or not. 

The rest of the paper is structured as follows; Section II describes the related work on the HSS topic. Section III describes our approach
on gathering our dataset. Section IV describes the machine learning methods and algorithms we used. In Section V we perform an extensive evaluation of the proposed framework and in Section VI we conclude the paper.

\section{Related Work}
\label{relatedwork}
As we mentioned, Hit Song Prediction is an active topic in MIR. Raza and  Nanath \cite{9190613} concluded  there is no magic formula yet
that could predict a song being hit before it is released. Various approaches have been introduced. Li-Chia Yang et al. \cite{7952230} introduced state-of-the-art deep learning techniques to the audio-based hit song prediction problem. Zangerle et al. \cite{Zangerle2019HitSP}  proposed a combination of low- and high-level audio features of songs in a deep neural network that distinguishes low- and high-level features to account for their particularities. Elena Georgieva et al. \cite{Georgieva2018HITPREDICTP} used five machine-learning algorithms and managed to predict the Billboard success of a song with approximately 75\% accuracy. Middlebrook and Sheik \cite{Middlebrook2019SongHP} tested four machine-learning models to achieve 88\% accuracy on predicting the Billboard success.

\section{Dataset and Features}
\label{dataset}

\begin{table}[!t]
\caption{Dataset's Features Specifications}
\begin{tabular}{|m{0.25\columnwidth}|m{0.65\columnwidth}|}
\hline
\rowcolor [gray]{0.7}Feature & Specification \\
\hline\hline
id & the song’s unique Spotify track ID\\
\hline
artist & the song's artist name\\
\hline
popularity & a value between 0 and 100, with 100 being the most popular\\
\hline
explicit &  a boolean value indicated whether a track has explicit lyrics\\
\hline
album type & the type of the album: one of “album”, “single”, or “compilation”\\
\hline
danceability & a value from 0.0 to 1.0 describing how suitable a track is for dancing\\
\hline
energy & a value from 0.0 to 1.0 that represents a perceptual measure of intensity and activity\\
\hline
key & the music key the track is in\\
\hline
loudness & the overall loudness of a track in decibels (dB)
\\
\hline
mode & indicates the modality (major=1 or minor=0) of a track\\
\hline
speechiness & a value from 0.0 to 1.0 describing the amount of spoken words present in the track\\
\hline
acousticness & a value from 0.0 to 1.0 predicting whether the track is acoustic\\
\hline
instrumentalness & a value from 0.0 to 1.0 predicting whether the track is instrumental or contains vocals\\
\hline
liveness & a value from 0.0 to 1.0 that describes the presence of an audience in the track\\
\hline
valence & a value from 0.0 to 1.0 describing the musical positiveness conveyed by a track\\
\hline
tempo & the overall estimated tempo of a track in beats per minute (BPM)\\
\hline
duration\_ms & the duration of the track in milliseconds\\
\hline
time\_signature & an estimated overall time signature of a track\\
\hline
\end{tabular}
\label{tab:features}
\end{table}

In this section, insights related to the data collection and processing are discussed. In particular, descriptions of the tools and methods that are utilized for data collection are included, while, at the same time, the process of selecting and extracting the features of the sample is presented. Furthermore, techniques for dataset preparation, such as normalization and augmentation, are utilized and described in detail below.

\subsection{Acquiring the Data}
% hit dataset 861
Initially, we acquired data that is mandatory for the dataset preparation and consists of the Billboard top 100 hits for every year, starting from 2011 until 2021. This data is retrieved through the Billboard API using the Python's library \emph{billboard.py}\footnote{https://github.com/guoguo12/billboard-charts}.
% Phase 1: collect billboard hits
In particular, we construct a collection composed of 1000 hit songs (that were in some year's Billboard top 100), where each one of them is determined by its title and the corresponding artist, or artists in several cases. Afterwards, depending on the latter collection, we perform HTTP requests to the Spotify API, using the \emph{spotipy}\footnote{https://spotipy.readthedocs.io/en/2.19.0/} Python library, in order to retrieve information related to the 1000 Billboard Hits. In response for each request, Spotify API provided a set of 10 related objects, in JSON format. The "track id", the "artist id" and the song's "popularity" were included, among others, in every object as basic song's features. So, the object with the highest popularity was retrieved to be under consideration for the construction of the dataset.
% Phase 2: Collect random tracks
Afterwards, we populate the dataset with random songs from Spotify. To achieve this, we generate queries of random characters as track titles. These queries was then posted as HTTP requests to the Spotify API. Respectively, 50 songs were included in the response message, while we choose randomly 10 of those. In total, 2000 of these requests were posted, leading to 20,000 collected tracks from the Spotify. As mentioned for the hit songs, the same applies for the non-hits case about the basic features that came included in the response message as a JSON object.

% Extract more features
Considering the aforementioned data, we proceeded to an enrichment in terms of the corresponding features for the respective tracks. Towards this direction, the \emph{audio\_features} functionality of the \emph{spotipy} has been utilized. For each acquired track object and based on the \emph{id} feature, which is unique for every object, we perform the corresponding API Call to retrieve further features. Table \ref{tab:features} summarizes the extracted features for each track from Spotify, containing the corresponding specification. The collected data determines the basis for the creation of the dataset that will be used by the classifiers. However, further processing is required to create a proper dataset for this purpose.

\begin{figure}[h]
    \centering
    \includegraphics[width=0.9\columnwidth]{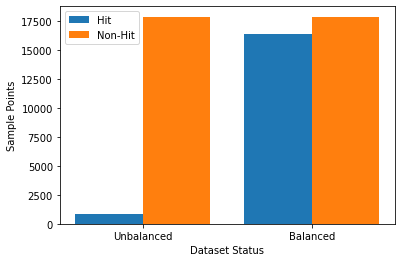}
    \caption{Unbalanced and Balanced Dataset}
    \label{fig:balancing}
\end{figure}

\subsection{Dataset Preparation}
% Balancing
To begin with, we proceeded with a data cleanup, in order to take into account only unique tracks from Spotify. Also, track objects with no data for specific features were discarded. After this initial process, a sample of approximately 18,000 Spotify songs occurred, where 861 of them were BillBoard Top 100 sometime between 2011 and 2021. Obviously, this sample instance is unbalanced, as there is no balanced ratio between the distribution of the classes "hit" and "non-hit". Data imbalances can affect classification predictions, when they are not managed properly. Thus, balancing the Dataset is mandatory before proceeding. One way to fight this issue is to generate new samples in the class which is under-represented. The most naive strategy is to generate new samples by randomly sampling with replacement the current available samples. The \emph{RandomOverSampler} offers such scheme and we utilized this technique from the sklearn python library\footnote{https://scikit-learn.org/stable/}.

% Scaling
Another issue that needs to be solved is that the audio features Spotify provides are not in the same scale. As Table \ref{tab:features} shows some features have boolean values, some have decibel values, some have a value between 0 and 100 and some between 0 and 1.0. Using scaling methods from the sklearn python library we managed to scale all numerical and boolean values between 0 and 1.0.

% feature importance
\begin{table}[!h]
\centering
\caption{PCA and Features Co-efficiency}
\begin{tabular}{|c|c|c|c|}
\hline
\rowcolor [gray]{0.7}Principal Component &  Most Variance Feature & Co-Efficiency\\
\hline\hline
PC1 & explicit & 0.728131 \\
\hline
PC2 & mode & -0.729190 \\
\hline
PC3 & key & -0.900311 \\
\hline
PC4 & acousticness & 0.651091 \\
\hline
PC5 & valence & -0.810128 \\
\hline
PC6 & danceability & -0.546361 \\
\hline
PC7 & popularity & -0.713593 \\
\hline
PC8 & tempo & -0.834668 \\
\hline
PC9 & instrumentalness & 0.711965 \\
\hline
PC10 & liveness & 0.604598 \\
\hline\hline
\hline
PC11 & duration\_ms & 0.044111 \\
\hline
PC12 & energy & -0.454230 \\
\hline
PC13 & loudness & -0.139564 \\
\hline
PC14 & speechiness & 0.239800 \\
\hline
PC15 & time\_signature & -0.050520 \\
\hline
\end{tabular}
\label{tab:pca}
\end{table}

After standardization, we utilized Principal Component Analysis, in order to identify the features that reflect to more information regarding to our classification problem, while reducing dataset's dimensions. It is critical to perform standardization prior to PCA, because the latter is quite sensitive regarding the variances of the initial variables. That is, if there are large differences between the ranges of initial variables, those variables with larger ranges will dominate over those with small ranges, leading to biased results. Principal components are new variables that are constructed as linear combinations of the initial variables. In our 15-dimensional dataset, 15 principal components are occurred, as these shown in Tab. \ref{tab:pca}, which also includes the most variance feature of each component. 

\begin{figure}[!ht]
    \centering
    \includegraphics[width=0.9\columnwidth]{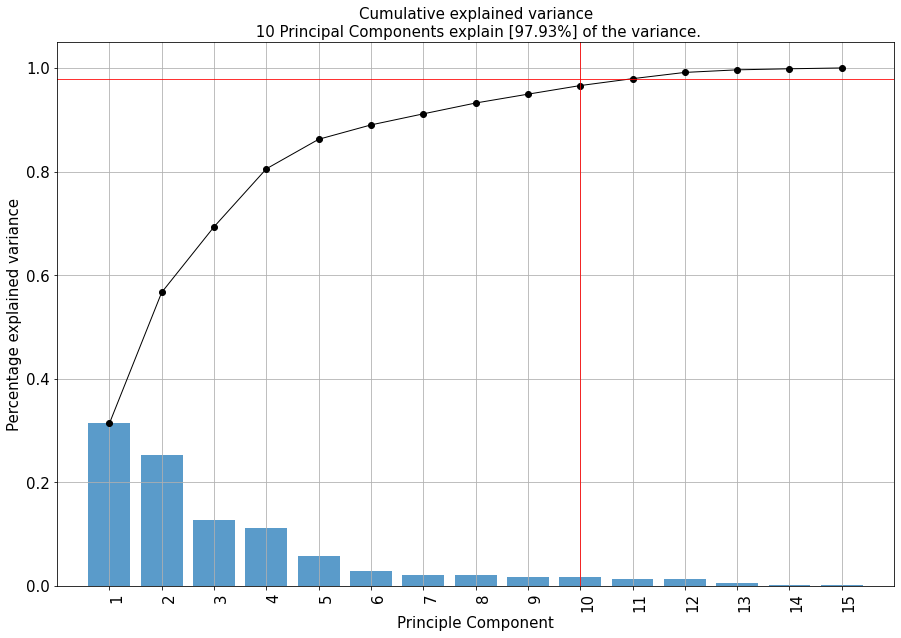}
    \caption{PCA Cumulative Explained Variance}
    \label{fig:pca}
\end{figure}

Features combinations are determined in such a way that the new ones (principal components) are uncorrelated and most of the information within the initial features is compressed into the first components.
Figure \ref{fig:pca} illustrates the variance against principal components. As it shown the first 10 components  contribute to significant variance, approximately 98\%, so the rest of them (11-15) can be ignored.

% Describe your dataset: how many training/validation/test examples do you have? Is there
% any preprocessing you did? What about normalization or data augmentation? What is the
% resolution of your images? How is your time-series data discretized? Include a citation on
% where you obtained your dataset from. Depending on available space, show some examples
% from your dataset. You should also talk about the features you used. If you extracted
% features using Fourier transforms, word2vec, histogram of oriented gradients (HOG), PCA,
% ICA, etc. make sure to talk about it. Try to include examples of your data in the report
% (e.g. include an image, show a waveform, etc.).

\section{Methods}
In this section different machine learning methods which have been used in this paper will be presented as follows: (i) \textit{Random Forest (RF)}, (ii) \textit{Support Vector Machine (SVM)}, (iii) \textit{Logistic Regression (LR)}, (iv) \textit{k-Nearest Neighbors (kNN)}
% \begin{itemize}
%     \item[-] Random Forest (RF)
%     \item[-] Support Vector Machine (SVM)
%     \item[-] Logistic Regression (LR)
%     \item[-] k-Nearest Neighbors (kNN)
% \end{itemize}

\subsection{Random Forest}
Random Forest (RF) models \cite{ho1995random} are one of the most popular methods in classification and aims to overcome the problem of decision trees, the over-fitting. In more details, they involve a set of generalized classification trees which are trained with different aspects of the dataset and randomly selected data. Its purpose is to decrease the overall variance of the remaining training data. In other words, this method tries to produce a homogeneous subset of the primary dataset by binary splitting it sequentially. RF models’ advantages are the high accuracy, the simplicity in training and the robustness against outliers. In their drawbacks belongs the fact that the function delivered is often discrete rather than smooth. The predictions for unseen samples y can be defined after training the model by averaging the predictions from all the individual classification trees g\textsubscript{b} on y, as it seems below.

\begin{equation}
f=\frac {1} {B} {sum_{b=1}^{B} {g_b(y)}}
\end{equation}

\subsection{Support Vector Machine}
The support-vector machine (SVM) \cite{cortes1995support} was initially created to solve logistic or classification problems. This method generalizes the Random Forest one and its purpose is to find the most optimal hyper-plane that divide the data into two recognizable and well-defined classes. When the data which are used in training are sparse, then the SVM can effectively improve its quality to fit but simultaneously increases the computational demands. The Gaussian Radial Basis Function (RBF) is commonly used as the algorithm's kernel: 

\begin{equation}
e^{-g \left\Vert(x-y)^2\right\Vert},
\end{equation}

where g is a kernel parameter.

\subsection{Logistic Regression}
Logistic Regression (LR) \cite{cox1958regression} is a commonly used classification algorithm when the target variable is categorical. This method aims to create a correlation between features and the desirable output. Generally, exist two types of logistic regression problems, the binary and the multi-class ones. Logistic regression model uses a logistic function to determine a binary dependent variable in its main form. This function re-frames the log-odds to meaningful probabilities and finally the labeled values are "0" and "1". The logistic function is described by the equation

\begin{equation}
P(x)=\frac {1} {1 + {e^\frac{-(x-m)}{s}}}
\end{equation}

where m is a location parameter and s is a scale one respectively.

\subsection{k-Nearest Neighbors}
The k-Nearest Neighbors algorithm \cite{9065747} is an non-parametric classification algorithm and refers to supervised problems. Subsequently, it is applied to labeled data, which are then categorized into many classes. This algorithm is commonly used as a classifier and is widely known for its simplicity and effectiveness. Nevertheless, it can also be used in regression problems. In the first case the algorithm uses a metric, such as Hamming distance, to determine the class in which the object belongs while in the latter it uses Euclidean distance usually. Finally, the value of the parameter k is important for its performance since it affects the boundaries between the classes. Generally, there is not a strict rule for it and it depends on the data. In binary classification problem the output can be calculated as the class with the highest frequency from the K-most similar instances. For example the probability of class 0 is estimated as follows.

\begin{equation}
P(class=0)=\frac {count(class=0)} {count(class=0) + count(class=1)}
\end{equation}

\section{Experiments/Results/Discussion}

% Results 

\begin{table}[t]
\centering
\caption{Model Metrics on Validation Set}
\begin{tabular}{l c c c}
\hline
\hline
\\[-1em]
\multirow{2}{*}{\textbf{Classifiers}} & \multicolumn{3}{c}{\textbf{Metrics}} \\
\cline{2-4}
\\[-1em]
& \textbf{Accuracy} & \textbf{Precision} & \textbf{Recall}\\
\\[-1em]
Support Vector Machine & 0.73 & 0.70 & 0.83\\

\rowcolor[gray]{0.9}Logistic Regression & 0.66 & 0.64 & 0.76\\

\textbf{Random Forest} & \textbf{0.86} & \textbf{0.82} & \textbf{0.94}\\

\rowcolor[gray]{0.9} k-Nearest Neighbors (n=25) & 0.76 & 0.71 & 0.90\\

Support Vector Machine - opt & 0.83 & 0.79 & 0.92\\

\rowcolor[gray]{0.9}k-Nearest Neighbors (n=25) - opt & 0.80 & 0.71 & 0.90\\
\\[-0.8em]
\\[-0.8em]
\hline
\hline
\end{tabular}
\label{tab:results 1}
\end{table}

\begin{figure*}[ht]
     \centering
     \begin{subfigure}[b]{0.32\textwidth}
         \centering
\includegraphics[width=\columnwidth]{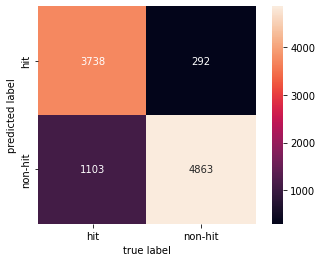}
         \caption{Random Forest Classifier}
         \label{fig:rf_cm}
     \end{subfigure}
     \hfill
     \begin{subfigure}[b]{0.32\textwidth}
         \centering
         \includegraphics[width=\columnwidth]{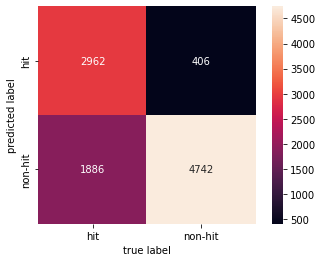}
         \caption{Optimized kNN Classifier}
         \label{fig:optsvm_cm}
     \end{subfigure}
     \hfill
     \begin{subfigure}[b]{0.32\textwidth}
         \centering
         \includegraphics[width=\columnwidth]{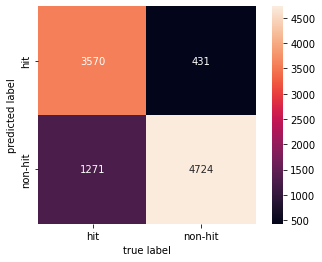}
         \caption{Optimized SVM Classifier}
         \label{fig:optsvm_cm}
     \end{subfigure}
     \caption{Confusion Matrices For Validation Set}
     \label{fig:CMs_valid}
\end{figure*}

\begin{figure*}[ht]
     \centering
     \begin{subfigure}[b]{0.32\textwidth}
         \centering
\includegraphics[width=\columnwidth]{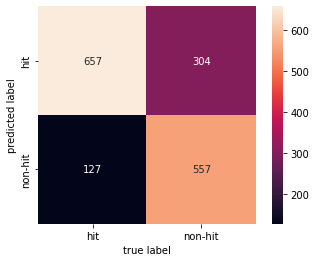}
         \caption{Random Forest Classifier}
         \label{fig:rf_cm}
     \end{subfigure}
     \hfill
     \begin{subfigure}[b]{0.32\textwidth}
         \centering
         \includegraphics[width=\columnwidth]{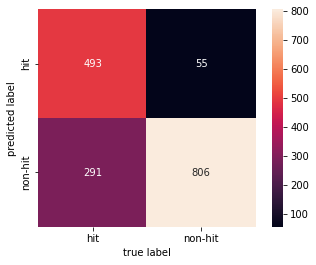}
         \caption{Optimized kNN Classifier}
         \label{fig:optsvm_cm}
     \end{subfigure}
     \hfill
     \begin{subfigure}[b]{0.32\textwidth}
         \centering
         \includegraphics[width=\columnwidth]{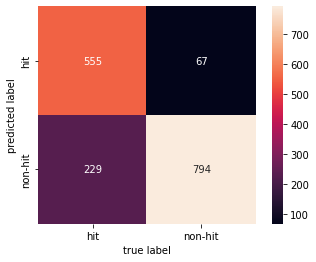}
         \caption{Optimized SVM Classifier}
         \label{fig:optsvm_cm}
     \end{subfigure}
     \caption{Confusion Matrices For Test Set}
     \label{fig:CMs_small}
\end{figure*}

% \begin{figure}[!ht]
%     \centering
%     \includegraphics[width=\columnwidth]{figs/cm_rf.png}
%     \caption{Random Forest Confusion Matrix}
%     \label{fig:fig1}
% \end{figure}

% You should also give details about what (hyper)parameters you chose (e.g. why did you
% use X learning rate for gradient descent, what was your mini-batch size and why) and how
% you chose them. Did you do cross-validation, if so, how many folds? Before you list your
% results, make sure to list and explain what your primary metrics are: accuracy, precision,
% AUC, etc. Provide equations for the metrics if necessary. For results, you want to have a
% mixture of tables and plots. If you are solving a classification problem, you should include a
% confusion matrix or AUC/AUPRC curves. Include performance metrics such as precision,
% recall, and accuracy. For regression problems, state the average error. You should have
% both quantitative and qualitative results. To reiterate, you must have both quantitative and
% qualitative results! This includes also unsupervised learning. Include visualizations of
% results, heatmaps, examples of where your algorithm failed and a discussion of why certain
% algorithms failed or succeeded. In addition, explain whether youthink you have overfit to
% your training set and what, if anything, you did to mitigate that.
% Make sure to discuss the figures/tables in your main text throughout this section. Your
% plots should include legends, axis labels, and have font sizes that are legible when printed.

% validation results (raw models and metrics' definition)
In this section we present the training process alongside an extensive evaluation of the machine learning models as those have been described on section IV. 

Firstly, we created the train, validation, and test sets, depending on the final version of the dataset, which essentially refers to scaled data with balanced sample point for each class. The train and validation sets were produced by splitting our balanced dataset in the ratio 7:3. So, approximately 25,000 music tracks were used, in order to train the discussed models. The test set was created by combining 861 unique hit and 861, randomly chosen, unique non-hit songs. 

For the evaluation, we use three different metrics; accuracy, precision and recall\footnote{https://blog.paperspace.com/deep-learning-metrics-precision-recall-accuracy/}. 

Accuracy is a metric that generally describes how the model performs across all classes. It is calculated as the ratio between the \textit{number of correct predictions} to the \textit{total number of predictions}: 

\[Accuracy=\frac{True{_{positive}}+True{_{negative}}}{Total \ number \ of \ predictions}\]
The precision is calculated as the ratio between the \textit{number of positive samples correctly classified} to the \textit{total number of samples classified as positive (either correctly or incorrectly)}:

\[Precision=\frac{True{_{positive}}}{True{_{positive}}+False{_{positive}}}\]
The recall is calculated as the ratio between the \textit{number of positive samples correctly classified as positive} to the \textit{total number of positive samples}:
\[Recall=\frac{True{_{positive}}}{True{_{positive}}+False{_{negative}}}\]

The results on our validation test are shown on Table \ref{tab:results 1}. We observe that the best performing classifiers are the Random Forest, the optimized SVM and the optimzed kNN. In an effort to achieve high performance on the kNN algorithm, we used cross-validation and found that the optimal value of neighbors is 25. Regarding the SVM model, the regularization parameter is set equal to 10, with an RBF kernel.
Confusion matrices for the performance of these classifiers on our validation test are shown on Figure \ref{fig:CMs_valid}. % optimization
In order to achieve better results, we used pipeline methods on the SVM and kNN classifiers. As Table \ref{tab:results 1} shows, using pipelines increased all metrics of these two algorithms by an average of 9\%.
% test set 

\begin{figure}[h]
    \centering
    \includegraphics[width=0.9\columnwidth]{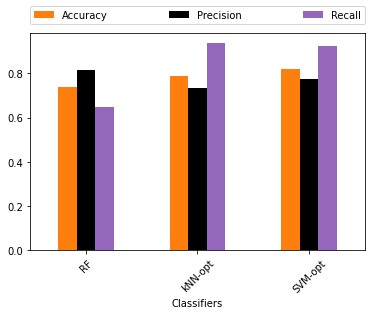}
    \caption{Model Metrics for Test Set}
    \label{fig:test_metrics}
\end{figure}

Aiming to evaluate the adaptability of our methods, we tested the three best performing classifiers on our test set. Result metrics are shown on Figure \ref{fig:test_metrics}, while confusion matrices are shown on Figure \ref{fig:CMs_small}. We see that Random Forest outperforms SVM and kNN on precision, while the other two used models display high recall values and better accuracy.
% discussion

Considering that music trends change constantly throughout the years \cite{interiano}, the Hit Song Prediction problem is of high complexity. On our study we used hit songs from the last decade in order to focus on contemporary music trends. 

High recall values on our models mean that we have a low number of false negatives. This is due to the fact that audio features collected from Spotify offer an interpretation on how a song sounds, without regard to the music trends of the year it was released. Thus, training models using these features lead to more accurate predictions about a song being non-hit than a song becoming a hit. Improving our models' precision, meaning that our models will predict better if a song will become a hit, requires implementing features that take under consideration the current music trends. 

Therefore, the results' footprint can be summarized in these questions: \\

\emph{(i) Would the \#1 hit song of 2011 be a hit if it was released in 2021?} \\

\emph{(ii) The prediction of the \#1 hit song of 2011 being a hit in 2021 implies over-fitting?}\\

\emph{(iii) Could a non-hit song of 2011 be a hit in 2021 if its features match the dominant features of hit songs of 2021?}

\section{Conclusion \& Future Work}

% Summarize your report and reiterate key points. Which algorithms were the highestperforming? Why do you think that some algorithms worked better than others? For
% future work, if you had more time, more team members, or more computational resources,
% what would you explore?

Our study showed the highest performing ML-algorithm for Hit Song Prediction was Random Forest having achieved 86\% accuracy. Random Forest achieved high precision on both valid and test set, making it suitable for the Hit Song Prediction problem. SVM and kNN algorithms illustrated higher accuracy on our test set, showing that they may be more effective in more extended experiments and therefore should be further explored and optimized.

Furthermore, exploring and gathering more metadata about songs would be useful in order to select better features. Having access to information about the songs' melody and harmony should give a better understanding about a song's structure, leading to better results.
Additionally, features like mood or emotion could also be considered during training.

Driven by the questions discussed on section V, exploring parameters and defining features that express music trends is another task that will be useful on approaching the Hit Song Prediction problem.

\section*{Acknowledgment}
This study is carried out in the context of the machine learning course of the Data Science and Machine Learning master degree program of School of Electrical and Computer Engineering of National Technical University of Athens. 

We thank Giannis Delimpaltadakis (Postdoctoral Researcher at Eindhoven University of Technology) for useful discussions and wish him the best for the next stage of his academic career.

% \section*{References}

\bibliography{conference_101719}{}
\bibliographystyle{IEEEtran}
% \begin{thebibliography}
% \end{thebibliography}

\end{document}